# Strong enhancement of critical current density in $MgB_2$ superconductor using carbohydrate doping

J. H. Kim, S. Zhou, M. S. A. Hossain, A. V. Pan & S. X. Dou*

Institute for Superconducting and Electronic Materials, University of Wollongong, Northfields Ave., NSW 2522, Australia

*e-mail: shi@uow.edu.au

With the relatively high critical temperature ($T_c$) of 39 K[1] and the high critical current density ($J_c$) of $> 10^5$ Acm$^{-2}$ in moderate fields, magnesium diboride ($MgB_2$) superconductors could offer the promise of important large-scale and electronic device applications to be operated at 20 K. A significant enhancement in the electromagnetic properties of $MgB_2$ has been achieved through doping with various form of carbon (C)[2-13]. However, doping effect has been limited by the agglomeration of nano-sized dopants and the poor reactivity of C containing dopants with $MgB_2$. Un-reacted dopants result in a reduction of superconductor volume. In this work, we demonstrate the advantages of carbohydrate doping over other dopants, resulting in an increase of in-field $J_c$ by more than one order of magnitude without any degradation of self-field $J_c$. As there are numerous carbohydrates readily available this finding has significant ramifications not only for the fabrication of $MgB_2$ but also for many C based compounds and composites.

Many of the important applications of superconductors require the maintenance of high $J_c$ in strong magnetic fields. $MgB_2$ has been fabricated in various forms: single crystals, bulk, thin films, tapes and wires. Various novel techniques have been directed towards the fabrication of technically usable high-$J_c$ $MgB_2$ wires[2,8,10,11]. The performance of $MgB_2$ can rival and exceed that of conventional superconductors. To take advantage of its $T_c$ of 39K, enhancements of the both upper critical field ($H_{c2}$) and $J_c$ are essential. Attempts to accomplish this have invoked the introduction of numerous techniques including chemical doping[2-14], irradiation[15], and various thermo-mechanical processing techniques[16-19]. Chemical doping is a simple and readily scalable technique. Since $MgB_2$ has a relatively large coherence length and small anisotropy, the fluxoids to be pinned are string-like and amenable to pinning by inclusions and precipitates in the grains. This opens a window to the success of chemical doping in this material.

Among the numerous forms of C-containing dopants, SiC doping has achieved a record high in-field $J_c(B)$, $H_{c2}$, and irreversibility ($H_{irr}$) in $MgB_2$[2,9,14]. These record high properties have been confirmed and reproduced by many groups[8-11], and the performance records remain unbroken up to now. However, the best high field $J_c$ values achieved in the SiC doped $MgB_2$ wires were compromised by the reduction in self-field and low-field $J_c$. Although nanosize precursor particles were chosen for the doping process it is a great challenge to achieve homogenous distribution of a small amount of nano-dopants within the matrix materials through solid state mixing. There are always agglomerates of nano-additives in the precursors. For various forms of C doping, the substitution of C for boron (B) can not be achieved at the same temperatures as that of the $MgB_2$ formation reaction due to their poor reactivity. Hydrocarbons such as benzene have been used for C substitution[20]. However, the C substitution level is limited by the high volatility of these compounds.

In order to overcome these problems we proposed to use a carbohydrate such as DL-malic acid ($C_4H_6O_5$) as the dopant. The significant advantages of carbohydrate doping are: (1) Carbohydrates can be dissolved in a solvent so that the solution can form a slurry with B powder. After evaporating the solvent the carbohydrate forms a coating on the B powder surfaces, giving a highly uniform mixture. (2) The carbohydrates in the mixture melt at lower temperatures and decompose at temperatures below the formation temperature of $MgB_2$, hence producing highly reactive and fresh C on the atomic scale, as well as a reducing reagent, carbon monoxide, which may convert boron oxide to B, reducing the impurities in B powder. (3) Because of the high reactivity of the freshly formed C, the C substitution for B can take place at the same temperature as the formation temperature of $MgB_2$. The simultaneous dual reactions promote C substitution for B in the lattice and the inclusion of excess C within the grains, resulting in the enhancement of $J_c$, $H_{irr}$, and $H_{c2}$.

In this study, we used malic acid as a representative of carbohydrate dopant. We fabricated $MgB_2$ superconductor with malic acid doping included. The lattice parameters, critical temperature ($T_c$), $J_c$, $H_{irr}$, $H_{c2}$, and microstructures are presented in comparison with the un-doped reference $MgB_2$. $MgB_2$ pellets were prepared by an *in-situ* reaction process with the addition of malic acid, $C_4H_6O_5$. The selected amount of $C_4H_6O_5$ (99%), from 0 to 30wt% of total $MgB_2$, was dissolved in toluene ($C_7H_8$, 99.5%). The solution was mixed with an appropriate amount of B (99%) powder. This slurry was dried in vacuum so that the B powder particles were coated by the $C_4H_6O_5$. This uniform composite was then mixed with an appropriate amount of Mg (99%) powder. These mixed powders were ground, pressed, and then sintered at 900°C for 30 min under high purity argon gas. The heating rate was 5°C/min. All samples were characterized by X-ray diffraction (XRD) and field emission gun scanning electron microscopy (FEG-SEM). The crystal structure was refined with the aid of the program FullProf. $T_c$ was defined as the onset temperature at which diamagnetic properties were observed. In addition, $H_{c2}$ and $H_{irr}$ were defined as $H_{c2}=0.9R(T_c)$ and $H_{c2}=0.1R(T_c)$ from the resistance (*R*) versus temperature (*T*) curve. The magnetization was measured at 5 and 20 K using a Physical Property Measurement System (PPMS, Quantum Design) in a time-varying magnetic field with sweep rate 50 Oe/s and amplitude 8.5 T. Since there is a large sample size effect on the magnetic $J_c$ for $MgB_2$[22,23] all the samples for measurement were made to the same size (1 x 2.2 x 3.3 mm$^3$) for comparison. The magnetic $J_c$ was derived from the width of the magnetization loop using Bean's model. $J_c$ versus magnetic field was measured up to 8.5 T.

Table 1 shows the measured data for the un-doped $MgB_2$ and $MgB_2+C_4H_6O_5$ samples with different addition levels. The lattice parameters calculated from XRD show a large decrease in the *a*-axis parameter with 10wt% $C_4H_6O_5$ and a small further drop in *a* with increasing $C_4H_6O_5$ addition level, but no change in the *c*-axis parameter. This is an indication of the C substitution for B. The actual C substitution level can be estimated from the *a*-axis change[21]. It should be noted that the net C percentage addition is only 36% of the $C_4H_6O_5$ addition. The actual C substitution levels of 1.9at% to 2.3at% of B at three doping levels are clearly higher than those with other forms of C dopants, which is attributable to the high reactivity of fresh C released from the decomposition of $C_4H_6O_5$ at low temperature (~150°C). The increase in sintering temperature improves both the crystallinity and the C substitution for B. The former will increase $T_c$, while the latter will decrease $T_c$. As a compromise, these two opposing factors result in a high level of C substitution for B with a relatively small drop in $T_c$. The high-field $J_c$s of the $MgB_2+C_4H_6O_5$ samples were much higher than that of the un-doped $MgB_2$. Specifically, it should be noted that the self-field $J_c$ of $MgB_2+C_4H_6O_5$ samples was not reduced at doping levels as high as 30wt% $C_4H_6O_5$, hence the connectivity between $MgB_2$ grains was not affected by doping with $C_4H_6O_5$. Although there is a possibility of the formation of $H_2O$ during sintering due to the decomposition of $C_4H_6O_5$, there was no degradation in self-field $J_c$, even for 30wt% $C_4H_6O_5$ added to $MgB_2$. This may be attributable to the fact that the decomposition products, C and CO, of $C_4H_6O_5$ reduced $B_2O_3$ and hence increased the effective cross section of the superconductor.

Figure 1(a) shows the magnetic field dependence of $J_c$ in all samples at 20 K and 5 K. It should be noted that $J_c$ values in high field were increased by more than an order of magnitude. For example, the $J_c$ value of $2.5 \times 10^4$ Acm$^{-2}$ at 5 K and 8 T for $MgB_2+30$wt% $C_4H_6O_5$ sample is higher than that of the un-doped $MgB_2$ by a factor of 21. In addition, there was no $J_c$ degradation in self-field for the $MgB_2+30$wt% $C_4H_6O_5$ sample. These findings can be further supported by the flux pinning results. Fig 1(b) plots the field dependence of the volume pinning force, $F_p = J \times B$, of all samples at 20 K. The $F_p$ is normalized by the maximum volume pinning force, $F_{p,max}$. The flux pinning for the $MgB_2+C_4H_6O_5$ samples was significantly higher than that of the un-doped one at $B>1.5$ T. This result indicates that the $F_p(B)$ of $MgB_2+C_4H_6O_5$ samples was improved by the C substitution effect and nano-C inclusions within the grains.

The normalized temperature dependence of $H_{irr}$ and $H_{c2}$ for all samples is shown in Figure 2. Significantly enhanced $H_{irr}$ and $H_{c2}$ for $MgB_2+C_4H_6O_5$ samples were observed, suggesting that C substitution into B sites results in an enhancement in $H_{irr}$ and $H_{c2}$. The steeper slopes of $H_{irr}$ for $MgB_2+C_4H_6O_5$ samples exceeded $H_{c2}$ of un-doped $MgB_2$ below a temperature of 22 K. The resistivity, $\rho$, for the un-doped and $MgB_2+C_4H_6O_5$ samples is 34 and 80-90 μΩ·cm at 40 K, respectively, as shown in Table 1. The increased resistivity for $MgB_2+C_4H_6O_5$ samples indicates the increased impurity scattering as a result of C substitution into B sites.

FEG-SEM images for (a) un-doped $MgB_2$, (b) $MgB_2$+10wt% $C_4H_6O_5$, and (c) $MgB_2$+30wt% $C_4H_6O_5$ are shown in Figure 3. The un-doped $MgB_2$ sample appears inhomogeneous, consisting of crystalline grains from several tens of nm in size to 500nm. The morphology of the $MgB_2$+10wt% $C_4H_6O_5$ sample was refined to smaller, denser, and more homogeneous grains compared to the un-doped $MgB_2$ one. The grain refinement by 10 wt% and 20 wt% $C_4H_6O_5$ doping is supported by the FWHM results for all the peaks as shown in Fig. 4. The small $MgB_2$ grains are effective in enhancing flux pinning because the grain boundaries of $MgB_2$ act as effective pinning centers, as in the case of A15 metallic superconductors. As the doping level further increases to 30wt%, however, grains appear to have a bar/plate shape, with their width up to 150 nm and length up to 400 nm, in a well connected grain network (Figure 3(c)). Consistent with the FEG-SEM image is the decrease in FWHM for the 30 wt% doped sample (Figure 4) although the average FWHM values for all peaks are still bigger than those of the un-doped sample. The FEG-SEM image suggests that at higher doping levels $C_4H_6O_5$ may act as a sintering aid to improve the crystallinity. The grain growth should not improve the electromagnetic properties. However, this effect may be offset by the increase in C substitution level, the reduction in resistivity (Table 1), and improvement in grain connectivity. This is well evidenced by the fact that the self-field $J_c$ of the $MgB_2$+30wt% $C_4H_6O_5$ sample was enhanced while the improved in-field $J_c$, $H_{irr}$, and $H_{c2}$ were maintained, as shown in Figures 1 and 2.

In summary, carbohydrate doping results in a small depression in $T_c$ but significantly increases the C substitution level, reduces the impurities, and hence improves $J_c$, $H_{irr}$, and $H_{c2}$ performance at all the operating temperatures and over the entire field range. Carbohydrates are cheap, abundant, and readily available. This finding opens a new direction for the manufacture of nano-doped materials using the carbohydrate solution route, which solves the agglomeration problem, avoids the use of expensive nano-additives, and achieves improved performance properties. As carbohydrates consist of a large range of materials, this work will have significant implications for further improvement of the performance properties of $MgB_2$, as well as many other C-based compounds and composites.

**Acknowledgements**
The authors gratefully acknowledge financial support from the Australian Research Council, Hyper Tech Research Inc, and CMS Alphatech International Ltd.

The authors declare no conflict financial interest in this work.
Corresponding author: S.X. Dou, shi@uow.edu.au

**Figure captions**

Table 1. Measured data for un-doped $MgB_2$ and $MgB_2+C_4H_6O_5$ samples with different addition levels.

Figure 1. Superconducting properties of un-doped $MgB_2$ and $MgB_2+C_4H_6O_5$ samples with different addition levels: (a) Magnetic field dependence of $J_c$ in all samples at 20 K and 5 K; (b) Field dependence of the volume pinning force, $F_p=J \times B$, of all samples at 20 K. The $F_p$ is normalized by the maximum volume pinning force, $F_{p,max}$.

Figure 2. Normalized temperature dependence of $H_{irr}$ and $H_{c2}$ for un-doped and $C_4H_6O_5$ doped samples. $H_{c2}$ and $H_{irr}$ were defined as $H_{c2}=0.9R(T_c)$ and $H_{c2}=0.1R(T_c)$ from the $R$ versus $T$ curve.

Figure 3. The photographs from field emission gun-scanning electron microscopy (FEG-SEM): (a) Un-doped $MgB_2$; (b) $MgB_2+10wt\% C_4H_6O_5$; (c) $MgB_2+30wt\% C_4H_6O_5$.

Figure 4. Full width at half maximum (FWHM) as a function of amount of $C_4H_6O_5$. $MgB_2$ (100), (101), (002), and (110) correspond to $2\theta \sim 33.6°$, $42.5°$, $52.0°$, and $60.0°$, respectively.

**Table 1.**

| DL-malic acid amount (wt%) | Lattice parameters | | Actual C (x) in MgB$_{2-x}$C$_x$ † | $T_c$ (K) | $\rho_{40\,K}$ (μΩ·cm) | $\rho_{300\,K}$ (μΩ·cm) | $H_{irr}$* (T) (20 K) | $J_c$ (Acm$^{-2}$) | |
|---|---|---|---|---|---|---|---|---|---|
| | a (Å) | c (Å) | | | | | | Self-field (20 K) | 8 T (5 K) |
| 0 | 3.08355 | 3.52175 | | 37.6 | 34.5 | 73.5 | 5.4 | 3.9 x 10$^5$ | 0.1 x 10$^4$ |
| 10 (3.6 wt% C) | 3.07516 | 3.52683 | 0.0380 | 35.8 | 90.2 | 146.5 | 6.7 | 3.5 x 10$^5$ | 2.3 x 10$^4$ |
| 20 (7.2 wt% C) | 3.07464 | 3.52297 | 0.0404 | 35.7 | 83.8 | 146.2 | 6.8 | 3.5 x 10$^5$ | 2.7 x 10$^4$ |
| 30 (10.8 wt% C) | 3.07319 | 3.52147 | 0.0460 | 35.8 | 79.6 | 131.9 | 6.7 | 4.0 x 10$^5$ | 2.6 x 10$^4$ |

† extrapolation from measured lattice parameters [20]

$H_{irr}$* was calculated from the standard criterion of critical current density (100 Acm$^{-2}$)

(a)

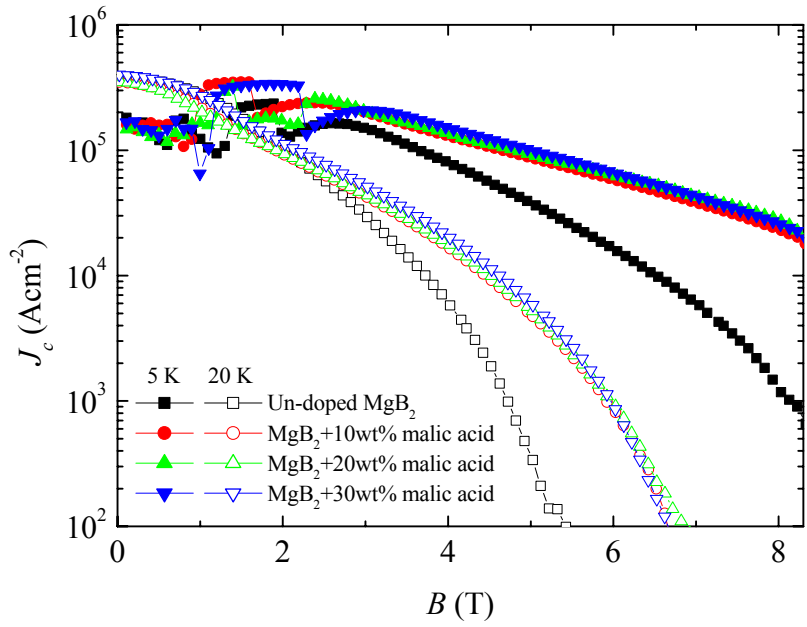

(b)

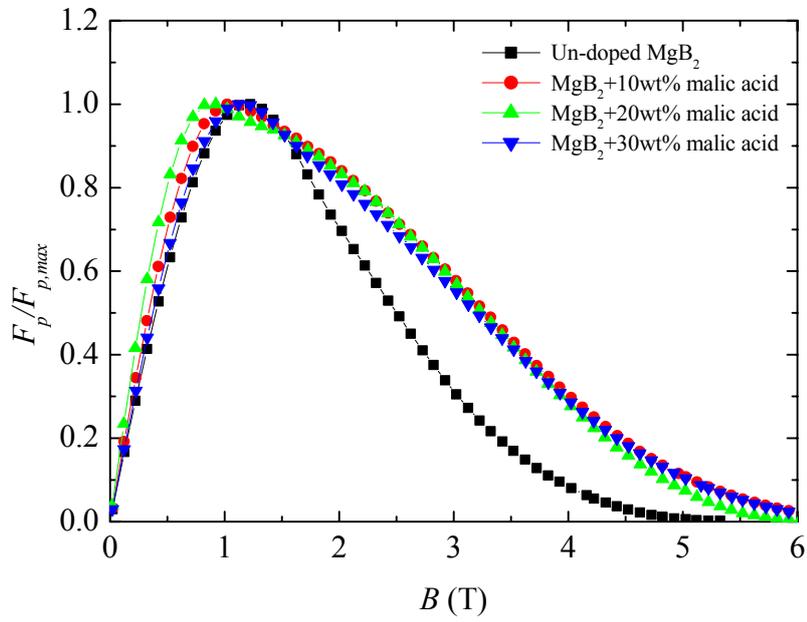

**Figure 1**

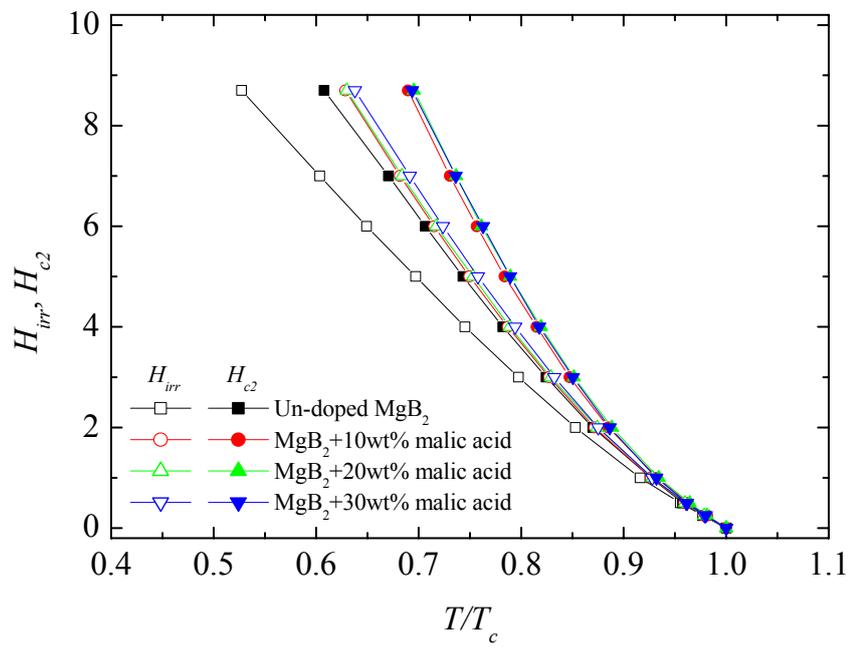

**Figure 2.**

(a)

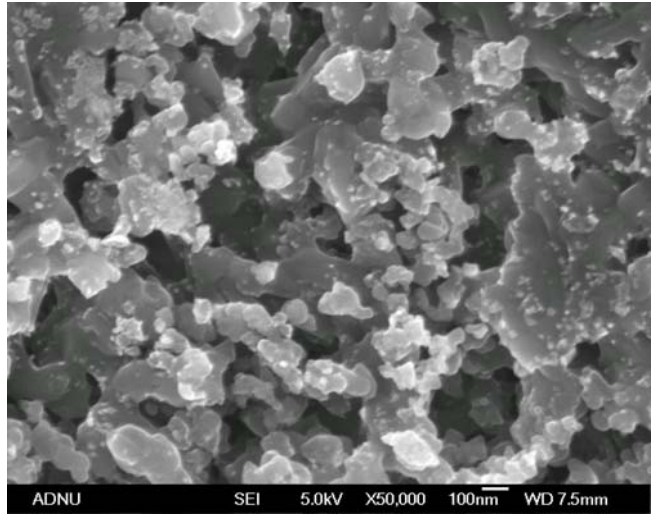

(b)

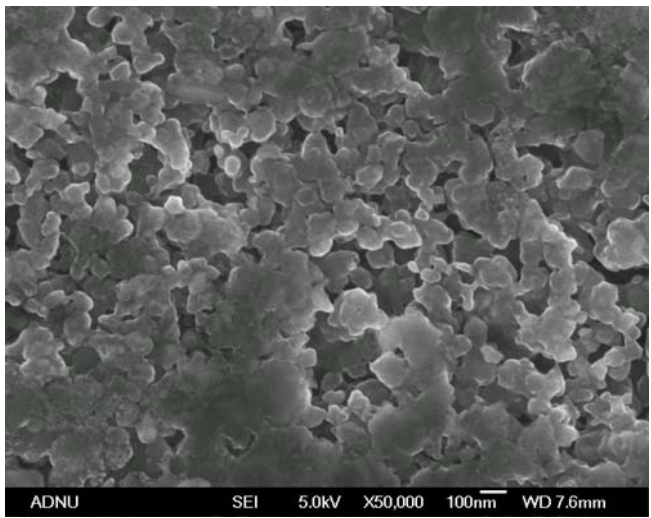

(c)

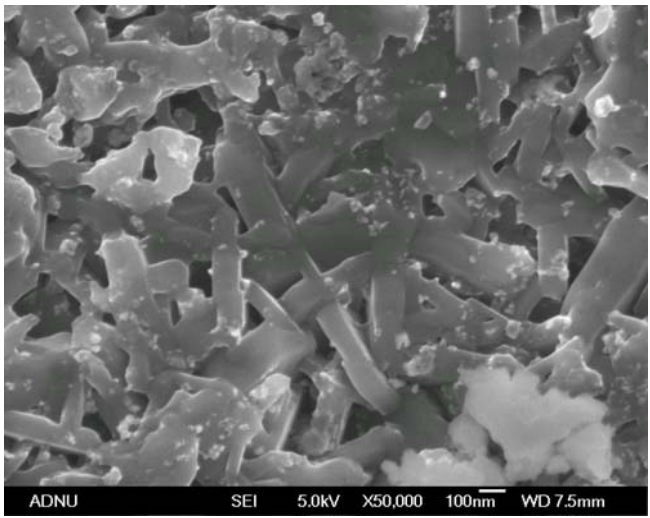

**Figure 3.**

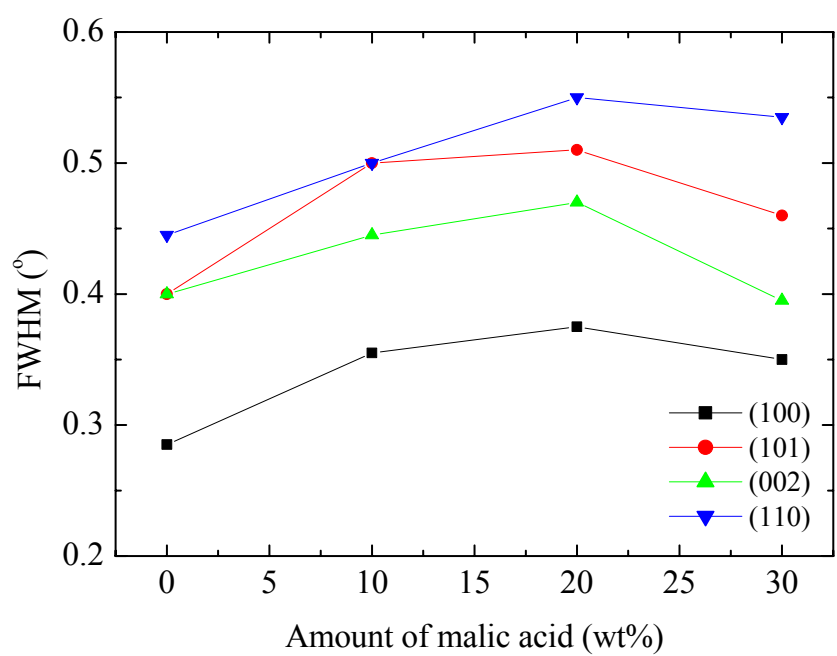

**Figure 4.**